\documentclass[12pt]{article}%
\usepackage{amsmath}
\usepackage{cite}
\usepackage{graphicx}
\usepackage{amsfonts}
\usepackage{amssymb}
\usepackage{appendix}%
\providecommand{\U}[1]{\protect\rule{.1in}{.1in}}

\csname @addtoreset\endcsname{equation}{section}
\newcommand{\eq}{\begin{equation}}
\newcommand{\feq}{\end{equation}}
\newcommand{\eqn}{\begin{eqnarray}}
\newcommand{\feqn}{\end{eqnarray}}
\newcommand{\arr}{\begin{eqnarray*}}
\newcommand{\farr}{\end{eqnarray*}}

\newcommand{\bea}{\begin{eqnarray}}
\newcommand{\eea}{\end{eqnarray}}

\begin{document}
\begin{titlepage}
\begin{flushright}
hep-th/yymmnnn\\
\end{flushright}
\vspace{.3cm}
\begin{center}
\renewcommand{\thefootnote}{\fnsymbol{footnote}}
{\Large{\bf Phantom Space-times  in Fake Supergravity}}
\vskip1cm
\vskip 1.3cm
Maryam Bu Taam$^1$  and Wafic A. Sabra$^2$
\vskip 1cm
{\small{\it
$^1$Physics Department, American University of Beirut\\
Beirut, Lebanon\\}}
\vskip .6cm {\small{\it
$^2$Centre for Advanced Mathematical Sciences and Physics Department\\
American University of Beirut\\ Lebanon  \\}}
\end{center}
\bigskip
\begin{center}
{\bf Abstract}
\end{center}
We discuss phantom metrics admitting Killing spinors in fake $N=2$, $D=4$ supergravity coupled to vector multiplets.
The Abelian $U(1)$ gauge fields in the fake theory have kinetic terms with the wrong sign. We solve the Killing spinor equations for the standard and fake theories in a unified fashion by introducing a
parameter which distinguishes between the two theories. The solutions found are  fully determined in terms of
algebraic conditions, the so-called stabilisation equations, in which the symplectic sections are related to a
set of functions. These functions are harmonic in the case of the standard supergravity theory and satisfy
the wave-equation in flat (2+1)-space-time in the fake theory.
Explicit examples are given for the minimal models with quadratic prepotentials.
\end{titlepage}

\section{Introduction}

In recent years, a good amount of research activity has been focused on the
classification of solutions preserving fractions of supersymmetry in
supergravity theories in various space-time dimensions. Finding new
gravitational solutions by solving first order Killing spinors differential
equations is certainly an easier task than solving for the coupled second
order Einstein equations of motion. Building on the work of Gibbons and Hull
\cite{hullgib}, Tod in \cite{Tod} performed the first systematic
classification for all metrics admitting Killing spinors in four-dimensional
Einstein-Maxwell theory. The solutions with time-like Killing spinors turn out
to be the known IWP solutions \cite{IWP} which in the static limit reduce to
the MP solutions \cite{mp}. More recently, techniques, partly based on
\cite{lawson}, were implemented in the classifications of supersymmetric
solutions. This was first done in \cite{first} and later has been a very
powerful tool in the classification of solutions in supergravity theories in
four and five space-time dimensions (see for example \cite{recentlower}). This
classification included, in addition to the standard ungauged and gauged
supergravities, fake de Sitter supergravity theories which can be obtained by
analytic continuation of anti de Sitter supergravity. It must be noted that de
Sitter supergravities can also be obtained by a non-linear Kaluza Klein
reduction of the so called * theories of Hull \cite{hull}. The reduction of
IIB* string theory and M* theory produced de Sitter supergravities with vector
multiplets in four and five space-time dimensions \cite{jimwen}. A new feature
about these theories is that they come with gauge fields with the
non-conventional sign of kinetic terms in the action. We shall refer to such
gauge fields as anti or phantom fields and gravitational solutions to such
theories as phantom solutions.

Phantom black hole solutions have been considered and analysed in
\cite{dyson}. Also, phantom solutions have been used by many authors in
astrophysics and in the field of dark matter (see for instance \cite{comref}
and references therein). In a recent work \cite{phantom1}, metrics with
space-like Killing vectors admitting Killing spinors in four-dimensional
Einstein gravity coupled to a phantom Maxwell field were found. These
solutions can be considered as the time-dependent analogues of the IWP metrics
of the canonical Einstein-Maxwell theory. While the IWP\ metrics are expressed
in terms of a harmonic complex function, the phantom analogue is expressed in
terms of a complex function satisfying the wave-equation in a flat $(2+1)$-space-time.

Generalisations of the IWP\ solutions in the context of $N=2$ supergravity
action coupled to matter multiplets were found sometime ago in \cite{BLS}.
These stationary solutions are generalisations of the double-extreme and
static black hole solutions found in \cite{sab}. In our present work, we will
generalise the results of \cite{phantom1} to four-dimensional $N=2$
supergravity theory coupled to vector multiplets. We shall consider the action%

\begin{equation}
\mathbf{e}^{-1}\mathcal{L}=\frac{1}{2}R-g_{A\bar{B}}\partial_{\mu}%
z^{A}\partial^{\mu}\bar{z}^{B}-\frac{\kappa^{2}}{4}\left(  \operatorname{Im}%
\mathcal{N}_{IJ}F^{I}\cdot F^{J}+\operatorname{Re}\mathcal{N}_{IJ}F^{I}%
\cdot\tilde{F}^{J}\right)  , \label{action}%
\end{equation}
where we used the notation $F^{I}\cdot F^{J}=F_{\mu\nu}^{I}F^{J\mu\nu}$and
$\tilde{F}^{J}=\ast$ $F^{J},$ $I=0,...,n.$ For $\kappa=i,$ this is the action
of the standard $N=2,$ $D=4$ supergravity theory coupled to vector multiplet.
For $\kappa=1,$ this represents the action of a fake theory where the gauge
field terms in the action come with the opposite sign. The $n$ complex scalar
fields, $z^{A},$ of $N=2$ vector multiplets are coordinates of a special
K\"{a}hler manifold and $g_{A\bar{B}}=\partial_{A}\partial_{\bar{B}}K$ is the
K\"{a}hler metric with $K$ being the K\"{a}hler potential. The structure of
the scalar fields and relations of special geometry remain unaltered in the
fake case. For details of special geometry, we refer the reader to
\cite{speone} and references therein.

In what follows, we give some details and relations of special geometry which
will be relevant to our discussions. A useful definition of a special
K\"{a}hler manifold can be given by introducing a ($2n+2)$-dimensional
symplectic bundle over the K\"{a}hler-Hodge manifold with the covariantly
holomorphic sections $\mathcal{V},$%

\begin{equation}
\mathcal{V}=\left(
\begin{array}
[c]{c}%
L^{I}\\
M_{I}%
\end{array}
\right)  =e^{K/2}\left(
\begin{array}
[c]{c}%
X^{I}\\
F_{I}%
\end{array}
\right)  ,\text{ \ \ \ }I=0,...,n,\text{ \ \ \ \ \ }\mathcal{D}_{\bar{A}%
}\mathcal{V}=0,
\end{equation}
where $\mathcal{D}_{\bar{A}}\mathcal{V}=\left(  \partial_{\bar{A}}-\frac{1}%
{2}\partial_{\bar{A}}K\right)  \mathcal{V}$ and $\mathcal{D}_{A}%
\mathcal{V}=\left(  \partial_{A}+\frac{1}{2}\partial_{A}K\right)  \mathcal{V}%
$. These sections obey the symplectic constraint%

\begin{equation}
i\left(  \bar{L}^{I}M_{I}-L^{I}\bar{M}_{I}\right)  =1. \label{symplectic}%
\end{equation}
The K\"{a}hler potential can be obtained from the holomorphic sections by
\begin{equation}
e^{-K}=i\left(  \bar{X}^{I}F_{I}-X^{I}\bar{F}_{I}\right)  .
\end{equation}
The coupling matrix, $\mathcal{N}_{IJ},$ can be defined by%

\begin{equation}
F_{I}(z)=\mathcal{N}_{IJ}X^{J}(z),\text{ \ \ \ \ }\mathcal{D}_{A}%
F_{I}(z)=\mathcal{\bar{N}}_{IJ}\mathcal{D}_{A}X^{I}(z).
\end{equation}
We also note the very useful relations%

\begin{align}
g^{A\bar{B}}\mathcal{D}_{A}L^{M}\mathcal{D}_{\bar{B}}\bar{L}^{I}  &
=-\frac{1}{2}\left(  \text{Im}\mathcal{N}\right)  ^{MI}-\bar{L}^{M}%
L^{I},\label{r}\\
F_{I}\partial_{\mu}X^{I}-X^{I}\partial_{\mu}F_{I}  &  =0. \label{rrr}%
\end{align}
Also, one can derive the relations \cite{para}%

\begin{align}
\mathcal{D}_{A}L^{I}dz^{A}  &  =\ \left(  d+i\mathcal{A}\right)
L^{I},\label{useful1}\\
dM_{I}-2\operatorname{Im}\mathcal{N}_{IJ}L^{J}\mathcal{A}  &  =\mathcal{\bar
{N}}_{IJ}dL^{J},\label{useful2}\\
\mathcal{A}  &  =M_{I}d\bar{L}^{I}-L^{I}d\bar{M}_{I}, \label{useful3}%
\end{align}
\ where the $U(1)$ K\"{a}hler connection $\mathcal{A}$ is defined by%

\begin{equation}
\mathcal{A}=-\frac{i}{2}(\partial_{A}Kdz^{A}-\partial_{\bar{A}}Kd\bar{z}^{A}).
\label{gc}%
\end{equation}
The Killing spinor equations we shall analyse are given by%

\begin{equation}
\left(  \nabla_{\mu}+\frac{i}{2}\mathcal{A}_{\mu}\gamma_{5}+\frac{\kappa}%
{4}\text{Im}\mathcal{N}_{IJ}\gamma\cdot F^{I}\left(  \operatorname{Im}%
L^{J}-i\gamma_{5}\operatorname{Re}L^{J}\right)  \gamma_{\mu}\right)
\varepsilon=0, \label{kilon}%
\end{equation}
and
\begin{equation}
\frac{\kappa}{2}(\text{Im}\,\mathcal{N})_{IJ}\gamma\cdot F^{J}\left[
\operatorname{Im}(g^{A\bar{B}}\mathcal{D}_{\bar{B}}{\bar{L}}^{I})-i\gamma
_{5}\operatorname{Re}(g^{A\bar{B}}\mathcal{D}_{\bar{B}}{\bar{L}}^{I})\right]
\varepsilon+\gamma^{\mu}\partial_{\mu}\left(  \operatorname{Re}z^{A}%
-i\gamma_{5}\operatorname{Im}z^{A}\right)  \varepsilon=0. \label{kiltw}%
\end{equation}

Here $\nabla_{\mu}=(\partial_{\mu}+\frac{1}{4}\gamma.\omega_{\mu})$ and
$\varepsilon$ are Dirac spinors. For $\kappa=i,$ those represent the vanishing
of the supersymmetry variations, in a bosonic background, of the gravitini and
gaugini in the standard $N=2,$ $D=4$ supergravity theory coupled to vector
multiplet. For $\kappa=1,$ those represent the vanishing of fake supersymmetry
transformations for a theory where all the gauge fields terms in the action
come with the opposite sign.

In our analysis of the Killing spinor equations, we follow the method of
spinorial geometry. We write the spinors as complexified forms on
$\mathbb{R}^{2}$. A generic spinor, $\varepsilon,$ can therefore be written as%

\begin{equation}
\varepsilon=\lambda1+\mu_{i}e^{i}+\sigma e^{12}, \label{gen}%
\end{equation}
where $e^{1}$, $e^{2}$ are 1-forms on $\mathbb{R}^{2}$, and $i=1,2$;
$e^{12}=e^{1}\wedge e^{2}$. $\lambda$, $\mu_{i}$ and $\sigma$ are complex functions.

The action of $\gamma$-matrices on these forms is given by%

\begin{align}
\gamma_{0} &  =-e^{2}\wedge+i_{e^{2}},\text{ \ \ \ \ \ \ \ }\gamma_{1}%
=e^{1}\wedge+i_{e^{1}},\nonumber\\
\gamma_{2} &  =e^{2}\wedge+i_{e^{2}},\text{ \ \ \ \ \ \ \ \ \ }\gamma
_{3}=i(e^{1}\wedge-i_{e^{1}}).
\end{align}
and $\gamma_{5}$ is defined by $\gamma_{5}=i\gamma_{0123}$ where%

\begin{equation}
\gamma_{5}1=1,\quad\gamma_{5}e^{12}=e^{12},\quad\gamma_{5}e^{i}=-e^{i}%
,\ \ \text{\ \ \ }i=1,\text{ }2. \label{fa}%
\end{equation}
Using the results of \cite{4spinor}, we define%

\begin{align}
\gamma_{+}  &  ={\frac{1}{\sqrt{2}}}(\gamma_{2}+\gamma_{0})=\sqrt{2}i_{e^{2}%
},\text{ \ \ \ \ \ }\nonumber\\
\gamma_{-}  &  ={\frac{1}{\sqrt{2}}}(\gamma_{2}-\gamma_{0})=\sqrt{2}%
e^{2}\wedge,\nonumber\\
\gamma_{1}  &  ={\frac{1}{\sqrt{2}}}(\gamma_{1}+i\gamma_{3})=\sqrt{2}i_{e^{1}%
},\text{ \ \ \ }\nonumber\\
\gamma_{\bar{1}}  &  ={\frac{1}{\sqrt{2}}}(\gamma_{1}-i\gamma_{3})=\sqrt
{2}e^{1}\wedge, \label{acts}%
\end{align}
where the non-vanishing metric components in this null basis are given by
$g_{+-}=1,g_{1\bar{1}}=1$. The canonical forms of the spinor are basically
representatives up to gauge transformations which preserve the Killing spinor
equation. Using $Spin(3,1)$ gauge transformations, it was shown in
\cite{4spinor}, that one finds the three canonical forms:
\begin{equation}
\text{\ \ \ }\varepsilon=1+\mu_{2}e^{2},\text{ \ \ }\varepsilon=1+\mu_{1}%
e^{1},\text{ \ \ \ }\varepsilon=e^{2}. \label{orbits}%
\end{equation}
As in \cite{phantom1}, we shall focus on the first canonical form$.$ Plugging
$\varepsilon=1+\mu e^{2}$ in (\ref{kilon}) and (\ref{kiltw}) and using
(\ref{acts}), the Killing spinor equations amount to two sets of equations:%

\begin{align}
\omega_{+,-1}  &  =0,\nonumber\\
\omega_{1,-1}  &  =0,\nonumber\\
\omega_{-,+1}  &  =0,\nonumber\\
\omega_{1,+1}  &  =0,\nonumber\\
\mu\omega_{-,-1}+i\kappa\sqrt{2}\text{Im}\,\mathcal{N}_{IJ}F_{-1}^{I}\bar
{L}^{J}  &  =0,\nonumber\\
\mu\omega_{\bar{1},-1}-\frac{i\kappa}{\sqrt{2}}\text{Im}\,\mathcal{N}%
_{IJ}\left(  F_{1\bar{1}}^{I}+F_{-+}^{I}\right)  \bar{L}^{J}  &
=0,\nonumber\\
\partial_{-}\log\mu-\frac{1}{2}\left(  \omega_{-,1\bar{1}}+\omega
_{-,-+}\right)  -\frac{i}{2}\mathcal{A}_{-}-i\frac{\kappa}{\mu\sqrt{2}%
}\text{Im}\,\mathcal{N}_{IJ}\left(  F_{1\bar{1}}^{I}+F_{-+}^{I}\right)
\bar{L}^{J}  &  =0,\nonumber\\
\partial_{1}\log\mu-\frac{1}{2}\left(  \omega_{1,1\bar{1}}+\omega
_{1,-+}\right)  -\frac{i}{2}\mathcal{A}_{1}  &  =0,\nonumber\\
\partial_{+}\log\mu-\frac{1}{2}\left(  \omega_{+,1\bar{1}}+\omega
_{+,-+}\right)  -\frac{i}{2}\mathcal{A}_{+}  &  =0,\nonumber\\
\omega_{1,-+}-\omega_{1,1\bar{1}}+i\mathcal{A}_{1}  &  =0,\nonumber\\
\omega_{-,-+}-\omega_{-,1\bar{1}}+i\mathcal{A}_{-}  &  =0,\nonumber\\
\partial_{\bar{1}}\log\mu-\frac{1}{2}\left(  \omega_{\bar{1},1\bar{1}}%
+\omega_{\bar{1},-+}\right)  -\frac{i}{2}\mathcal{A}_{\bar{1}}+\frac{i\kappa
}{\mu}\text{Im}\,\mathcal{N}_{IJ}F_{+\bar{1}}^{I}\bar{L}^{J}\sqrt{2}  &
=0,\nonumber\\
\frac{1}{2}\left(  \omega_{\bar{1},-+}-\omega_{\bar{1},1\bar{1}}%
+i\mathcal{A}_{\bar{1}}\right)  -i\kappa\mu\text{Im}\,\mathcal{N}_{IJ}%
F_{-\bar{1}}^{I}L^{J}\sqrt{2}  &  =0,\nonumber\\
\frac{1}{2}\left(  \omega_{+,-+}-\omega_{+,1\bar{1}}+i\mathcal{A}_{+}\right)
-i\frac{\kappa\mu}{\sqrt{2}}\text{Im}\,\mathcal{N}_{IJ}\left(  F_{-+}%
^{I}-F_{1\bar{1}}^{I}\right)  L^{J}  &  =0,\nonumber\\
\omega_{+,+1}-i\kappa\mu\text{Im}\,\mathcal{N}_{IJ}F_{+1}^{I}L^{J}\sqrt{2}  &
=0,\nonumber\\
\omega_{\bar{1},+1}+i\frac{\kappa\mu}{\sqrt{2}}\text{Im}\,\mathcal{N}%
_{IJ}\left(  F_{1\bar{1}}^{I}-F_{-+}^{I}\right)  L^{J}  &  =0, \label{setone}%
\end{align}
and%

\begin{align}
-i\kappa g^{A\bar{B}}\mathcal{D}_{\bar{B}}{\bar{L}}^{I}(\text{Im}%
\,\mathcal{N})_{IJ}\left(  F_{-+}^{J}-F_{1\bar{1}}^{J}\right)  +\partial
_{-}z^{A}\mu\sqrt{2}  &  =0,\nonumber\\
-i\bar{\kappa}\bar{\mu}g^{A\bar{B}}\mathcal{D}_{\bar{B}}{\bar{L}}%
^{I}(\text{Im}\,\mathcal{N})_{IJ}\left(  F_{1\bar{1}}^{J}-F_{-+}^{J}\right)
+\partial_{+}z^{A}\sqrt{2}  &  =0,\nonumber\\
2i\bar{\kappa}\bar{\mu}g^{A\bar{B}}\mathcal{D}_{\bar{B}}{\bar{L}}%
^{I}(\text{Im}\,\mathcal{N})_{IJ}F_{-\bar{1}}^{J}+\partial_{\bar{1}}z^{A}%
\sqrt{2}  &  =0,\nonumber\\
2i\kappa g^{A\bar{B}}\mathcal{D}_{\bar{B}}{\bar{L}}^{I}(\text{Im}%
\,\mathcal{N})_{IJ}F_{+1}^{J}+\partial_{1}z^{A}\mu\sqrt{2}  &  =0.
\label{set2}%
\end{align}
The analysis of the equations of (\ref{setone}) gives:%

\begin{align}
\operatorname{Im}\mathcal{N}_{IJ}F_{-\bar{1}}^{I}L^{J}  &  =-\frac
{i\bar{\kappa}}{\sqrt{2}|\mu|^{2}}\left(  \partial_{\bar{1}}+i\mathcal{A}%
_{\bar{1}}\right)  \bar{\mu},\nonumber\\
\text{Im}\mathcal{N}_{IJ}\left(  F_{-+}^{I}-F_{1\bar{1}}^{I}\right)  L^{J}  &
=i\kappa\sqrt{2}\left(  \partial_{-}+i\mathcal{A}_{-}\right)  \bar{\mu
},\nonumber\\
\text{Im}\,\mathcal{N}_{IJ}F_{+1}^{I}L^{J}  &  =-\frac{i\kappa}{\sqrt{2}%
}\left(  \partial_{1}+i\mathcal{A}_{1}\right)  \bar{\mu}, \label{gcon}%
\end{align}
with the condition%

\begin{equation}
\mu\partial_{-}\bar{\mu}+\kappa^{2}\partial_{+}\log\bar{\mu}=-i\left(
\mathcal{A}_{-}|\mu|^{2}+\kappa^{2}\mathcal{A}_{+}\right)  . \label{con}%
\end{equation}
We also obtain the following relations for the spin connection%

\begin{align}
\text{\ \ \ }\omega_{1\bar{1}}  &  =\left(  \partial_{+}\log\frac{\mu}%
{\bar{\mu}}-i\mathcal{A}_{+}\right)  \mathbf{e}^{+}+i\mathcal{A}_{-}%
\mathbf{e}^{-}+\partial_{1}\log\mu\mathbf{e}^{1}-\partial_{\bar{1}}\log
\bar{\mu}\mathbf{e}^{\bar{1}},\nonumber\\
\omega_{-1}  &  =\frac{\kappa^{2}}{|\mu|^{2}}\left(  \partial_{1}\log
\mu-i\mathcal{A}_{1}\right)  \mathbf{e}^{-}+\left(  \partial_{-}\log
\mu-i\mathcal{A}_{-}\right)  \mathbf{e}^{\bar{1}},\nonumber\\
\omega_{-+}  &  =\left(  \partial_{1}\log\mu-i\mathcal{A}_{1}\right)
\mathbf{e}^{1}+\left(  \partial_{\bar{1}}\log\bar{\mu}+i\mathcal{A}_{\bar{1}%
}\right)  \mathbf{e}^{\bar{1}}+\partial_{+}\log\bar{\mu}\mu\mathbf{e}%
^{+},\nonumber\\
\omega_{+1}  &  =\left(  \partial_{+}\log\bar{\mu}+i\mathcal{A}_{+}\right)
\mathbf{e}^{\bar{1}}+\kappa^{2}\mu\left(  \partial_{1}\bar{\mu}+i\mathcal{A}%
_{1}\bar{\mu}\right)  \mathbf{e}^{+}.
\end{align}
The vanishing of torsion implies the conditions%

\begin{equation}
d\mathbf{e}^{1}+d\log\bar{\mu}\wedge\mathbf{e}^{1}=0,
\end{equation}

\begin{equation}
d\mathbf{e}^{+}=-\left(  \partial_{-}\log\frac{\mu}{\bar{\mu}}-2i\mathcal{A}%
_{-}\right)  \mathbf{e}^{\bar{1}}\wedge\mathbf{e}^{1}-\left(  \frac{\kappa
^{2}}{|\mu|^{2}}\mathbf{e}^{-}-\mathbf{e}^{+}\right)  \wedge\left(  \left(
\partial_{\bar{1}}\log\bar{\mu}+i\mathcal{A}_{\bar{1}}\right)  \mathbf{e}%
^{\bar{1}}+\left(  \partial_{1}\log\mu-i\mathcal{A}_{1}\right)  \mathbf{e}%
^{1}\right)  ,
\end{equation}

and
\begin{align}
&  d\mathbf{e}^{-}=-\left(  \partial_{+}\log\frac{\bar{\mu}}{\mu
}+2i\mathcal{A}_{+}\right)  \mathbf{e}^{\bar{1}}\wedge\mathbf{e}^{1}%
+\partial_{+}\log|\mu|^{2}\mathbf{e}^{+}\wedge\mathbf{e}^{-}\nonumber\\
&  -\kappa^{2}\mathbf{e}^{+}\wedge\left(  \left(  \mu\partial_{1}\bar{\mu
}+i\mathcal{A}_{1}\mu\bar{\mu}\right)  \mathbf{e}^{1}+\left(  \bar{\mu
}\partial_{\bar{1}}\mu-i|\mu|^{2}\mathcal{A}_{\bar{1}}\right)  \mathbf{e}%
^{\bar{1}}\right) \nonumber\\
&  -\frac{1}{|\mu|^{2}}\mathbf{e}^{-}\wedge\left(  \left(  \bar{\mu}%
\partial_{1}\mu-i\mu\bar{\mu}\mathcal{A}_{1}\right)  \mathbf{e}^{1}+\left(
\mu\partial_{\bar{1}}\bar{\mu}+i|\mu|^{2}\mathcal{A}_{\bar{1}}\right)
\mathbf{e}^{\bar{1}}\right)  .
\end{align}
An immediate result of the torsion free conditions and (\ref{con}) is that
$\left(  \mu\bar{\mu}\mathbf{e}^{+}-\kappa^{2}\mathbf{e}^{-}\right)  $ is a
total differential%

\begin{equation}
d\left(  \mu\bar{\mu}\mathbf{e}^{+}-\kappa^{2}\mathbf{e}^{-}\right)  =0,
\end{equation}
and that the vector $V,$%

\begin{equation}
V=|\mu|^{2}\mathbf{e}^{+}+\kappa^{2}\mathbf{e}^{-}=|\mu|^{2}\partial
_{-}+\kappa^{2}\partial_{+}, \label{kilvec}%
\end{equation}
is a Killing vector which is space-like for $\kappa^{2}=1$ and time-like for
$\kappa^{2}=-1.$ Note that these two special vectors are related to the inner
Hermitian products $<\gamma_{0}\varepsilon,\gamma_{a}\varepsilon>$ and
$<\gamma_{0}\varepsilon,\gamma_{5}\gamma_{a}\varepsilon>.$

The above conditions enable us to introduce the coordinates $(t,x,y,z),$ such that%

\begin{align}
\mathbf{e}^{1} &  =\frac{1}{\bar{\mu}\sqrt{2}}\left(  dx+idy\right)
,\nonumber\\
\mathbf{e}^{+} &  =\frac{1}{|\mu|^{2}\sqrt{2}}\left(  dz+\kappa^{2}|\mu
|^{2}\left(  dt+\sigma\right)  \right)  ,\nonumber\\
\mathbf{e}^{-} &  =-\frac{\kappa^{2}}{\sqrt{2}}\left(  dz-\kappa^{2}|\mu
|^{2}\left(  dt+\sigma\right)  \right)  ,
\end{align}
and the metric is independent of the coordinate $t$ and is given by
\begin{equation}
ds^{2}=2\mathbf{e}^{1}\mathbf{e}^{\bar{1}}+2\mathbf{e}^{+}\mathbf{e}%
^{-}=\kappa^{2}|\mu|^{2}\left(  dt+\sigma\right)  ^{2}+\frac{1}{|\mu|^{2}%
}\left(  -\kappa^{2}dz^{2}+dx^{2}+dy^{2}\right)  .
\end{equation}
Here $\sigma$ is a one form, $\sigma=\sigma_{x}dx+\sigma_{y}dy+\sigma_{z}dz,$
independent of the coordinate $t$ and satisfies
\begin{equation}
d\sigma=-\frac{\kappa^{2}}{|\mu|^{2}}\ast_{3}\left(  id\log\frac{\mu}{\bar
{\mu}}+2\mathcal{A}\right)  ,\label{dsigma}%
\end{equation}
where $\ast_{3}$ is the Hodge dual with metric $\left(  -\kappa^{2}%
dz^{2}+dx^{2}+dy^{2}\right)  .$

The first two equations in second set of conditions (\ref{set2}) imply that%

\begin{equation}
\left(  \mu\bar{\mu}\partial_{-}+\kappa^{2}\partial_{+}\right)  z^{A}=0.
\label{con2}%
\end{equation}
Thus the scalar fields are also independent of the coordinate $t.$ Equations
(\ref{con2}) and (\ref{gc}) imply that%

\begin{equation}
\kappa^{2}\mathcal{A}_{+}+\mu\bar{\mu}\mathcal{A}_{-}=0.
\end{equation}
Going back to (\ref{con}), we then deduce that
\begin{equation}
\partial_{t}\mu=0.
\end{equation}
Multiplying the relations (\ref{set2}) by $\mathcal{D}_{A}L^{M}$ and using the
relations (\ref{r}) and (\ref{useful1}), we obtain the relations%

\begin{align}
\frac{i\kappa}{2}\left(  F_{-+}^{M}-F_{1\bar{1}}^{M}\right)  +i\kappa
(\text{Im}\,\mathcal{N})_{IJ}\left(  F_{-+}^{J}-F_{1\bar{1}}^{J}\right)
\bar{L}^{M}L^{I}+\left(  \mathcal{\partial}_{-}+i\mathcal{A}_{-}\right)
L^{M}\mu\sqrt{2}  &  =0,\nonumber\\
-2i\bar{\kappa}\bar{\mu}(\text{Im}\,\mathcal{N})_{IJ}F_{-\bar{1}}^{J}\bar
{L}^{M}L^{I}-i\bar{\kappa}F_{-\bar{1}}^{M}\bar{\mu}+\left(  \mathcal{\partial
}_{\bar{1}}+i\mathcal{A}_{\bar{1}}\right)  L^{M}\sqrt{2}  &  =0,\nonumber\\
-2i\kappa(\text{Im}\,\mathcal{N})_{IJ}F_{+1}^{J}\bar{L}^{M}L^{I}-i\kappa
F_{+1}^{M}+\left(  \mathcal{\partial}_{1}+i\mathcal{A}_{1}\right)  L^{M}%
\mu\sqrt{2}  &  =0,
\end{align}
which upon using (\ref{gcon}) and converting to space-time indices using%

\begin{equation}
\partial_{+}=\frac{|\mu|^{2}}{\sqrt{2}}\partial_{z},\text{ \ \ \ \ \ }%
\partial_{-}=-\frac{\kappa^{2}}{\sqrt{2}}\partial_{z},\text{ \ \ \ \ \ }%
\partial_{1}=\frac{\bar{\mu}}{\sqrt{2}}\left(  \partial_{x}-i\partial
_{y}\right)  ,\text{ \ }%
\end{equation}
we obtain for the gauge field strength two-form%

\begin{align}
F^{I}  &  =d\left(  i\kappa\mu L^{I}-i\bar{\kappa}\bar{L}^{I}\bar{\mu}\right)
\mathbf{\wedge}\left(  dt+\sigma\right)  -\frac{1}{|\mu|^{2}}\ast_{3}\left[
\kappa\bar{\mu}d\bar{L}^{I}-\kappa\bar{L}^{I}d\bar{\mu}+\bar{\kappa}\mu
dL^{I}-\bar{\kappa}L^{I}d\mu\right] \nonumber\\
&  -\frac{2i}{|\mu|^{2}}\ast_{3}\left(  \bar{\kappa}\mu L^{I}-\kappa\bar
{L}^{I}\bar{\mu}\right)  \mathcal{A}. \label{fs}%
\end{align}
Using (\ref{dsigma}), (\ref{fs}) can be rewritten in the form%

\begin{equation}
F^{I}=d\left[  \left(  i\kappa\mu L^{I}-i\bar{\kappa}\bar{L}^{I}\bar{\mu
}\right)  \left(  dt+\sigma\right)  \right]  -\ast_{3}d\left[  \kappa\left(
\frac{\bar{L}^{I}}{\mu}\right)  +\bar{\kappa}\left(  \frac{L^{I}}{\bar{\mu}%
}\right)  \right]  . \label{gf}%
\end{equation}
Calculating the dual $\tilde{F}^{I},$ we obtain%

\begin{align}
\tilde{F}^{I}  &  =\frac{i}{|\mu|^{2}}\ast_{3}d\left[  \kappa\bar{L}^{I}%
\bar{\mu}-\bar{\kappa}\mu L^{I}\right] \nonumber\\
&  +\left(  \left(  \bar{\kappa}\bar{L}^{I}d\bar{\mu}-\kappa\mu dL^{I}\right)
+\left(  \kappa L^{I}d\mu-\bar{\kappa}\bar{\mu}d\bar{L}^{I}\right)  \right)
\wedge\left(  dt+\sigma\right) \nonumber\\
&  -\left(  2i\mathcal{A}\left(  \kappa\mu L^{I}-\bar{\kappa}\bar{\mu}\bar
{L}^{I}\right)  \right)  \wedge\left(  dt+\sigma\right)  .
\end{align}
Again using (\ref{dsigma}) as well as (\ref{useful2}), we obtain%

\begin{equation}
\operatorname{Re}\mathcal{N}_{IJ}F^{J}-\operatorname{Im}\mathcal{N}_{IJ}%
\tilde{F}^{J}=d\left[  \left(  i\kappa\mu M_{I}-i\bar{\kappa}\bar{\mu}\bar
{M}_{I}\right)  \left(  dt+\sigma\right)  \right]  -\ast_{3}d\left[
\kappa\left(  \frac{\bar{M}_{I}}{\mu}\right)  +\bar{\kappa}\left(  \frac
{M_{I}}{\bar{\mu}}\right)  \right]  .
\end{equation}
Then Bianchi identities and Maxwell equations%

\begin{equation}
dF^{I}=0,\text{ \ \ \ \ \ \ \ \ \ \ \ }d\left(  \operatorname{Re}%
\mathcal{N}_{IJ}F^{J}-\operatorname{Im}\mathcal{N}_{IJ}\tilde{F}^{J}\right)
=0,
\end{equation}
imply, respectively, the conditions%

\begin{equation}
\left(  \frac{\kappa\bar{L}^{I}}{\mu}+\frac{\bar{\kappa}L^{I}}{\bar{\mu}%
}\right)  =\psi^{I},\text{ \ \ \ \ \ }\left(  \frac{\kappa\bar{M}_{I}}{\mu
}+\frac{\bar{\kappa}M_{I}}{\bar{\mu}}\right)  =\psi_{I},\label{stab}%
\end{equation}
where%
\begin{align}
\nabla^{2}\psi^{I} &  =\nabla^{2}\psi_{I}=0,\text{ \ \ \ \ \ \ \ }\nonumber\\
\nabla^{2} &  =\partial_{x}^{2}+\partial_{y}^{2}-\kappa^{2}\partial_{z}^{2}.
\end{align}
Using (\ref{stab}), (\ref{symplectic}), (\ref{rrr}) and (\ref{useful3}), we obtain%

\begin{equation}
\mathcal{A}=\frac{|\mu|^{2}}{2}\left(  \psi_{I}d\psi^{I}-\psi^{I}d\psi
_{I}\right)  -\frac{i}{2}d\log\frac{\mu}{\bar{\mu}}. \label{na}%
\end{equation}
Substituting (\ref{na}) back in the expression of $d\sigma$, we obtain%

\begin{equation}
d\sigma=-\kappa^{2}\ast_{3}\left(  \psi_{I}d\psi^{I}-\psi^{I}d\psi_{I}\right)
.
\end{equation}

For $\kappa=i,$ we obtain the known solutions of \cite{BLS,denef} which are
generalisations of the solutions first obtained in \cite{sab}. The new
derivation here, based on spinorial geometry, reveals that these are the
unique solutions with time-like Killing vector as has also been demonstrated
in \cite{ortin}. For $\kappa=1,$ we obtain new phantom solutions for theories
with the wrong signs for the gauge kinetic terms. In this case, the functions
$\psi^{I}$ and $\psi_{I}$ in (\ref{stab}) satisfy the wave-equation%

\begin{equation}
\left(  \partial_{x}^{2}+\partial_{y}^{2}\right)  \psi^{I}=\partial_{z}%
^{2}\psi^{I},\text{ \ \ \ }\left(  \partial_{x}^{2}+\partial_{y}^{2}\right)
\psi_{I}=\partial_{z}^{2}\psi_{I}.
\end{equation}
These solutions are the unique solutions with space-like Killing vectors
admitting Killing spinors.

\section{Examples: Quadratic Prepotentials}

Supergravity minimal models are characterised by quadratic prepotentials $F$
\cite{minimal}. For these models we have
\begin{equation}
M_{I}=\partial_{I}F=Q_{IJ}L^{J},
\end{equation}
where $Q_{IJ}$ is symmetric. Static black holes for the minimal models were
considered in \cite{minimal}. Without lack of generality, and as was explained
in \cite{minimal}, $Q_{IJ}$ can be taken to be purely imaginary. The
stabilisation conditions (\ref{stab}) for these models then give
\begin{equation}
\left[  \frac{\kappa\bar{L}^{I}}{\mu}+\frac{\bar{\kappa}L^{I}}{\bar{\mu}%
}\right]  =\psi^{I},\text{ \ \ \ \ \ }\left[  \frac{\bar{\kappa}L^{I}}%
{\bar{\mu}}-\frac{\kappa\bar{L}^{I}}{\mu}\right]  =Q^{IJ}\psi_{J}.
\end{equation}
This can be solved by%

\begin{equation}
L^{I}=\frac{\bar{\mu}}{2}\kappa\left(  \psi^{I}+Q^{IJ}\psi_{J}\right)
\text{.\ }%
\end{equation}
The symplectic constraint (\ref{symplectic}), then implies that%

\begin{equation}
\frac{1}{|\mu|^{2}}=\frac{i}{2}\left(  Q_{IJ}\psi^{J}\psi^{I}-Q^{IJ}\psi
_{I}\psi_{J}\right)  .
\end{equation}
Using (\ref{gf}), the gauge fields are given by%
\begin{equation}
F^{I}=d\left[  i|\mu|^{2}\kappa^{2}Q^{IJ}\psi_{J}\left(  dt+\sigma\right)
\right]  -\ast_{3}d\psi^{I}.
\end{equation}

For $\kappa=1,$ as in \cite{phantom1}, explicit solutions can be obtained if
one assumes that the solution depends on the coordinate $z$ only. In this case
we have%

\begin{equation}
\partial_{z}^{2}\psi^{I}=\partial_{z}^{2}\psi_{I}=0,
\end{equation}
and the solution can be given by
\begin{equation}
\psi^{I}=A^{I}+p^{I}z,\text{ \ \ \ }\psi_{I}=B_{I}+q_{I}z.
\end{equation}
For $A^{I}=B_{I}=0,$ the solution is then given by%

\begin{equation}
ds^{2}=\frac{\gamma^{2}}{z^{2}}\left(  dt\right)  ^{2}+\frac{z^{2}}{\gamma
^{2}}\left(  -dz^{2}+dx^{2}+dy^{2}\right)  \text{,}%
\end{equation}
where we have defined $\gamma^{-2}=\frac{i}{2}\left(  Q_{IK}p^{K}p^{I}%
-Q^{IM}q_{M}q_{I}\right)  .$ Setting%

\begin{equation}
\tau=\frac{z^{2}}{2\gamma},\text{ \ \ }x_{3}=\sqrt{\frac{\gamma}{2}}t,\text{
\ \ \ }x_{2}=\sqrt{\frac{2}{\gamma}}x,\text{ \ \ \ \ }x_{1}=\text{\ }%
\sqrt{\frac{2}{\gamma}}y,
\end{equation}
we get the Kasner metric%

\begin{equation}
ds^{2}=-d\tau^{2}+\tau\left(  dx_{2}\right)  ^{2}+\tau\left(  dx_{1}\right)
^{2}+\frac{1}{\tau}\left(  dx_{3}\right)  ^{2},
\end{equation}
where the gauge fields are given by%

\begin{equation}
F^{I}=\frac{1}{2}\gamma\left(  -i\frac{Q^{IJ}q_{J}}{\tau^{3/2}}d\tau\wedge
dx_{3}+p^{I}dx_{2}\wedge dx_{1}\right)  .
\end{equation}

\bigskip

In summary, we have obtained new phantom metrics admitting Killing spinors in
fake $N=2$, $D=4$ supergravity where the Abelian $U(1)$ gauge fields have
kinetic terms with the wrong sign. The solutions found are expressed in terms
of algebraic constraints satisfied by the symplectic sections. The solutions
are characterised in terms of a set of functions satisfying the wave-equation
in flat $(2+1)$-space-time. Explicit solutions are constructed for the
supergravity models where the prepotential is quadratic. Our analysis can be
generalised to fake gauged supergravity theories as well as to the de Sitter
supergravities constructed in \cite{jimwen}. Non-supersymmetric phantom
solutions can also be analysed using the general framework presented in
\cite{nonsusy}. We hope to report on this in a future publication.

\bigskip

\textbf{Acknowledgements}: The work of W. A. Sabra is supported in part by the
National Science Foundation under grant number PHY-1415659. M. Bu Taam would
like to thank the American University of Beirut for funding her Master's
degree during which this work took place.

\end{document}